\documentclass[11pt,twoside]{article}

\usepackage{asppreprint}
\usepackage{epsf}
\usepackage{psfig}
\usepackage{lscape}
\usepackage{natbib}

\begin{document}
\title{The Formation of Stellar Cusps in Galactic Nuclei}
\author{Brian W. Murphy} \affil{Department of Physics and Astronomy,
  Butler University, 4600 Sunset Ave, Indianapolis, IN 46208 USA}

\begin{abstract} 
The dynamics of galactic nuclei can be affected by several mechanisms.
Among these are stellar evolution, stellar collisions, mass
segregation, and tidal disruptions of stars due to the central black
hole. In this presentation I will address how each of these affects the
stellar cusp and the resulting observational signatures.  Using a
set of dynamically evolving Fokker-Planck simulations I present the
dynamical evolution a nuclear stellar cluster and the growth of the
central massive black hole within the Galactic Nucleus. In addition to
the Galactic Center I explore a wide variety of galactic nuclei and
their resulting stellar cusps.
\end{abstract}

\section{Introduction}  

Simulations by H\'{e}non (1961) were among the first to suggest that 
dense stellar systems may generate central stellar cusps, that is a
surface-brightness or stellar density that rises as a power-law into 
the cluster center.  One of the first hints of stellar cusps in dense
stellar systems was observed by King (1966).  His observations
indicated central brightness excesses in the most concentrated
globular clusters.  Wyller (1970) and Wolfe \& Burbidge (1970)
suggested that such excesses could be due to a central massive black
hole of several thousand solar masses that may have formed via a
collapse of the cluster core, what are currently considered
intermediate mass black holes. Shortly thereafter this central black
hole view was strengthened by the detection of X-ray sources in
globular clusters (Giacconi et al.~1974).  Bachall \& Ostriker (1975)
proposed that the source of these X-rays were accreting massive black
holes.  This view changed in 1980's with more precise positions of
X-ray sources measured with the \textsl{Einstein X-Ray Observatory}
(Grindlay et al.~1984).  These positions indicated that the sources
were only 1.5 $M_{\sun}$ and ruled out massive black holes as their
origin because of their distances from the cluster centers.  Today
we know that roughly one fifth of globular clusters show 
brightness excesses (Djorgovski \& King 1986).  Though there is some
indication that intermediate mass black holes could perhaps exist in
some globular clusters and galactic nuclei, it is now generally
thought that most of these cusps in clusters are due to a collapsed
stellar distribution of neutron stars, white dwarfs, and main sequence
stars mimicking the presence of a massive black hole (Baumgardt et
al.~2003).

Though most globular clusters are not thought to harbor massive black
holes, they have provided us with a laboratory to test stellar
dynamics and our knowledge of the formation of stellar cusps in dense
stellar systems.  Also the same computational methods applied to
globular clusters can easily be adapted to their larger mass cousins,
galactic nuclei. Young et al.~(1978) found such a cusp in the galaxy
M87. This was followed by yet more observations of not only our Galaxy
but other galaxies showing stellar cusps at their centers.

\section{Cusp Formation}

\subsection{Stellar Systems Without Central Black Holes}

One of the focal points in the dynamical evolution of globular
clusters has been the phenomenon of core collapse.  Core collapse is a
result of energy slowly being transferred via star-star scatterings
from the core of a cluster to its halo, causing the core to contract
and the halo to expand (Lightman \& Shapiro 1978).  In time this
contraction accelerates so that within a finite time the core will
collapse.  In the idealized identical star scenario this would take
roughly 15 half-mass relaxation times. Simulations of the core
collapse of globular clusters show that a cusp replaces the initially
flat core.  This stellar density cusp has the form of $\rho \propto
r^{-\beta}$ where $\beta$ has the value 2.2 for a system that is
composed of stars of a single mass (Cohn 1980).  This slope can vary
significantly though if stars are of differing masses.  Due to
equipartition of energy more massive objects will have lower
velocities, fall inward, and in time dominate the inner regions
of the cluster (Inagaki\& Saslaw 1985).  Because of this disparity in
mass the observed power-law can differ significantly from the actual
mass distribution within the cluster (Murphy \& Cohn 1988 and Grabhorn
et al.~1992).  In this multi-mass case the slope of the power-law
density profile is given by
\begin{equation}
\beta \simeq 1.9 {m_{rg}\over m_{rem}} + 0.3
\end{equation}
where $m_{rg}$ is the mass of the red giants and $m_{rem}$ is the mass
of the dominant remnants. 
\begin{figure}[!ht]
\epsscale{0.8}
\plotone{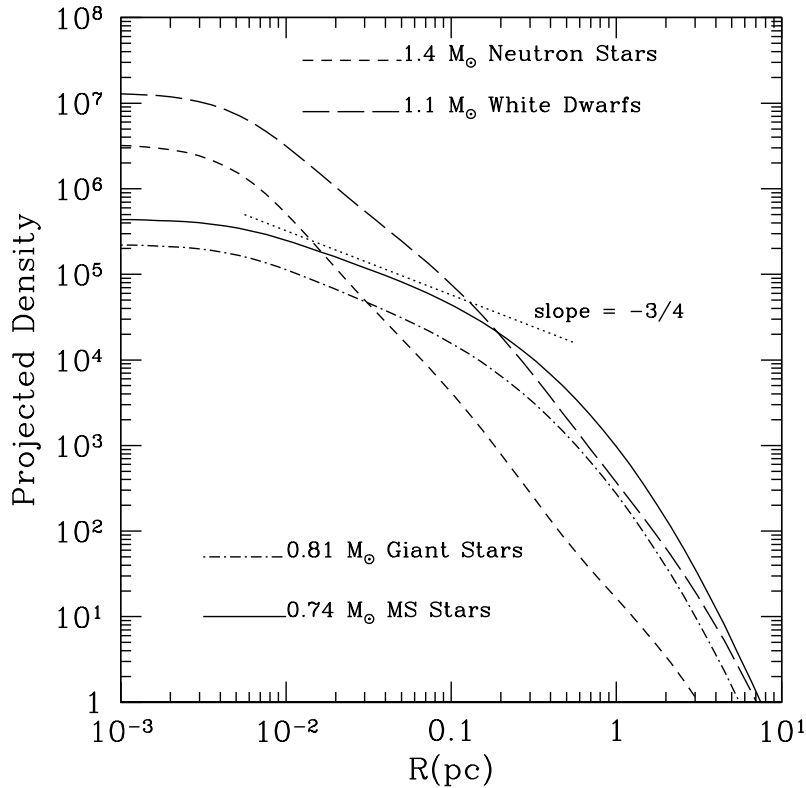} 
\caption{The projected stellar distribution of a globular cluster that
  has gone through core collapse.  Though no central black hole is
  included the power-law slope of the giants and main sequence stars
  mimic the cusp near a black hole.}
\end{figure}

Because neutron stars of 1.4 $M_{\odot}$ and white dwarfs over 1
$M_{\odot}$ are likely to be the most massive objects in globular
clusters they will dominate the inner regions of the cluster.  On the
other hand the observed red giants will be the most luminous stars and
have a mass close to 0.8 $M_{\odot}$.  This mass difference leads to a
$\beta$ near 7/4 which, as will be discussed later, is strikingly
similar to the power-law cusp found near a black hole.  Figure 1 shows
an example of such a cluster without central massive black hole. The
power-law slope of the red giant cusp flattens moving inward due to
massive white dwarfs and neutron stars dominating the innermost part
of the cluster.  For reference a -7/4 slope is indicated in Figure 1.
Roughly 20$\%$ of globular clusters appear to have undergone core
collapse (Djorgovski \& King 1986) and the most of these have
power-law slopes in projection close to -3/4.  The velocity-profile
will also be quite shallow in projection, $v \propto r^{-0.2}$.  It is
important to note that the presence of a stellar cusp need not be
caused by a massive black hole but can also be caused by mass
segregation.  To distinguish between the two possibilities other
parameters such as the velocity profile must be observed.  The
velocity profile in a Keplerian potential takes the form of $v \propto
r^{-0.5}$.

\subsection{Stellar Systems With Central Black Holes}

\begin{figure}[!ht]
\epsscale{0.90}
\plotone{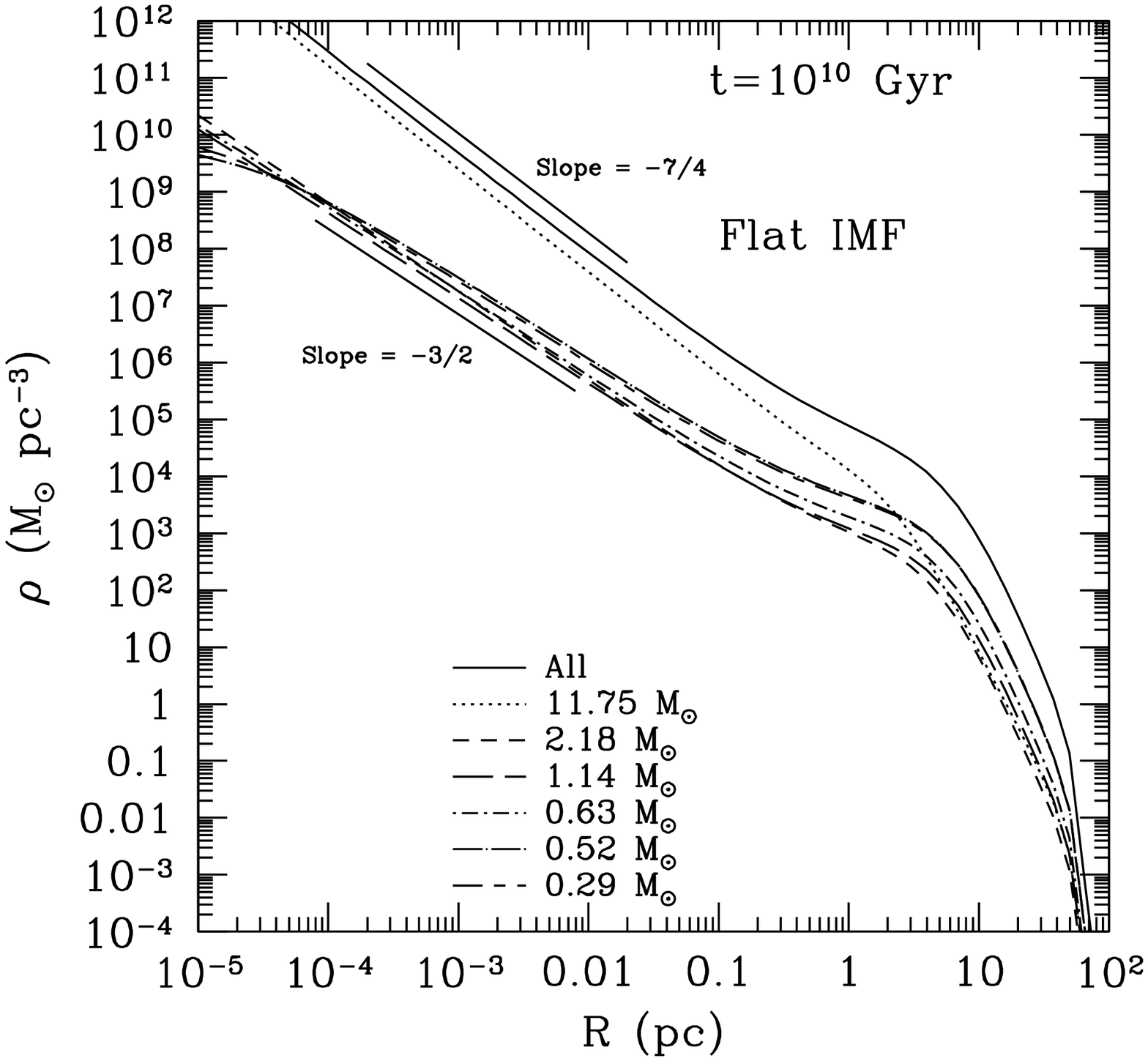} 
\vspace{-1.2cm} 
\caption{A simulation of a nucleus with a flat mass function. }
\end{figure}

\begin{figure}[!ht]
\epsscale{0.90}
\plotone{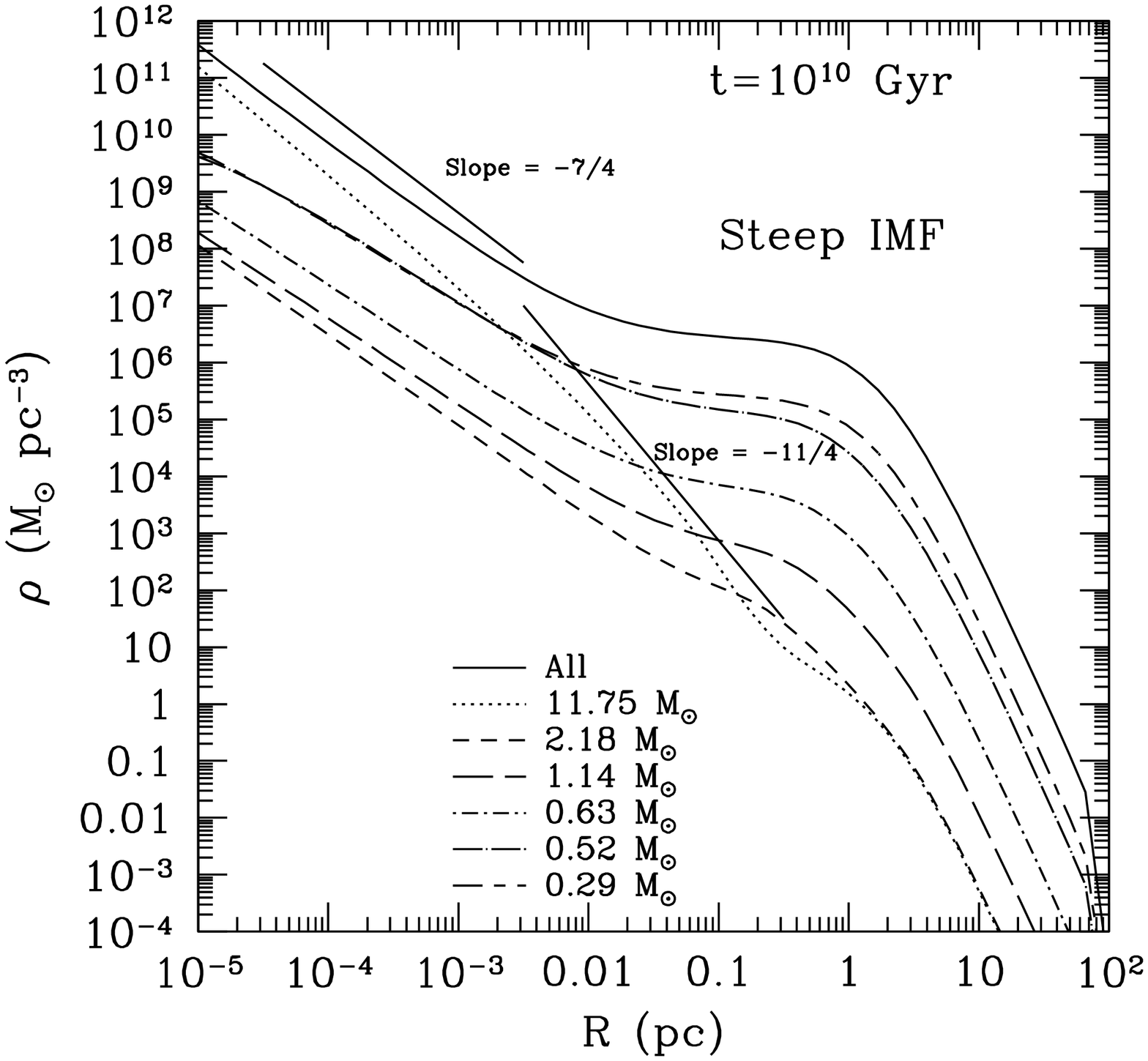}
\vspace{-1.2cm}
\caption{A simulation of a nucleus with a steep mass function. }
\end{figure}
In dense stellar systems with central massive black holes, the nature
of the resulting cusp will differ from that of globular clusters.  If
it is assumed that the massive black hole dominates the central
potential of the cluster then it will be Keplerian in nature.  Driven
by the hypothesis that globular clusters harbored massive black holes
one of the first attempts to determine the nature of cusps in
Keplerian potentials was Peebles (1972).  He found, using a semi
quantitative approach, that the signature of a central black hole
should be an observable cusp with a power-law form of $\rho\propto
r^{-9/4}$.  Later Bachall \& Wolf (1976) derived detailed expressions
for diffusion coefficients and numerically solved the Boltzmann
equation to find that the cusp should have the form $\rho\propto
r^{-7/4}$.  Bachall \& Wolf (1977) extended this to stellar systems
containing stars of two different masses.  In their models they found
that the power-law slope of the less massive component will be flatter
than the massive component. The resulting relation for the slope of the
stellar density cusp was given by
\begin{equation}
\beta = -\frac{d \ln \rho_i}{d \ln r} = - \left(\frac{m_i}{4m_1} + {3\over 2}
\right)
\end{equation}
From this relation there is a range of power-law slopes, -3/2 to -7/4
depending on the mass of stars dominating the stellar density profile.
This behavior is easily seen in the model shown in Figure 2.  This
model has a mass similar to that of the Galactic Nucleus but has a
relatively flat stellar mass function.  This mass function produces a
sizable number of stellar-mass black holes which by a Hubble time
dominate the inner 2 parsecs of the nucleus.  As would be expected
from equation 2 the black holes have a slope of -7/4 whereas the
lower-mass stars progressively approach the -3/2 limit.

At the other extreme of a steep stellar mass function Alexandar \&
Hopman (2009) have shown that when stellar-mass black holes do not
dominate the density profile their profile can be much steeper.  In
this case which they refer to as ``strong segregation'' the low-mass
main sequence stars will dominate the the stellar density profile and
have a-7/4 slope, causing the more massive stellar mass black holes to
have a power-law profile much steeper than -7/4.  In this case the
much less dominant but much more massive stellar-mass black holes have
a slope of -11/4.  An example of such a steep mass function is shown
in Figure 3.  Note that where the stellar-mass black holes do not
dominate they show this -11/4 slope.  But in the inner most regions,
$10^{-3}$ pc and inward, they dominate the density profile and revert
back to the Bachall \& Wolf -7/4 slope.

\subsection{The Effect of Stellar Collisions on the Cusp}

Stellar collisions can alter a cusp by removing stars from it.
Typically collisions remove stars from the high density, high velocity
regions of a nuclear cluster.  Therefore collisions have the largest
effect in the innermost regions of the nucleus.  Even though few stars
bound to the central black hole will remain in the most massive
nuclei, stars from larger radii, on high eccentricity orbits will still
populate the inner region of the nucleus.  In this scenario, collision
dominated regions of the cusp will have a much shallower profile,
typically showing a power-law slope of -1/2.  In projection this slope
would appear nearly flat.

\section{Realistic Models of Galactic Nuclei}

\subsection{The Galactic Nucleus}
\begin{figure}[!ht]
\epsscale{0.90}
\plotone{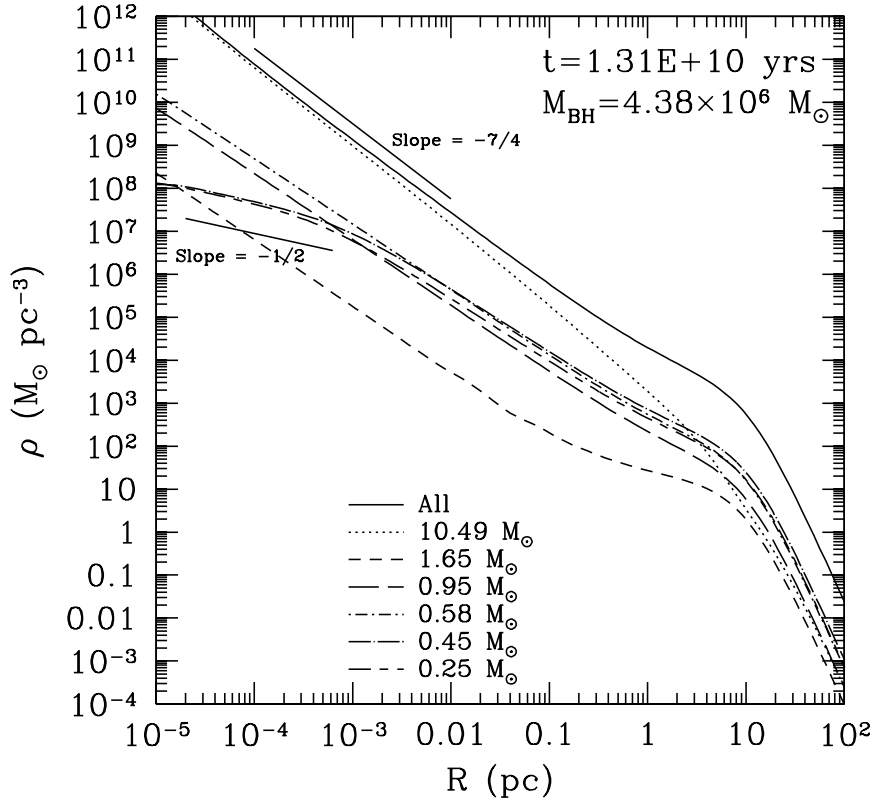}
\vspace{-1.12cm}
\caption{Stellar density profiles of a simulation of the Galactic Center.}
\end{figure}

\begin{figure}[!ht]
\epsscale{0.90}
\plotone{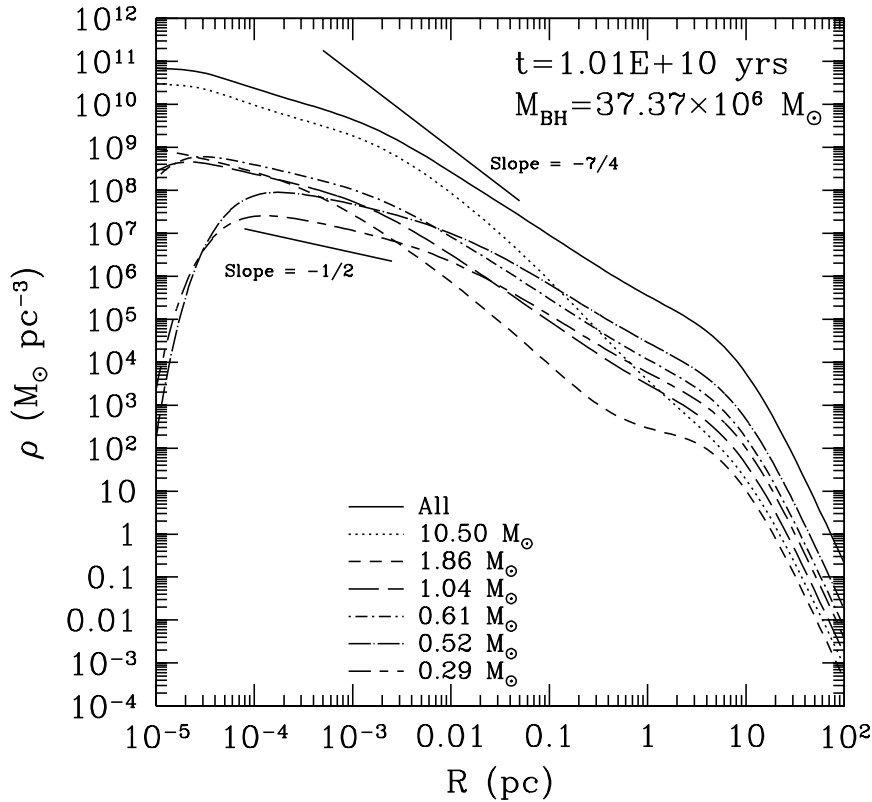} 
\vspace{-1.2cm}
\caption{Stellar density profiles of a simulation with 10 times the
  mass of the Galactic Center.}
\end{figure}

\begin{figure}[!ht]
\epsscale{0.90}
\plotone{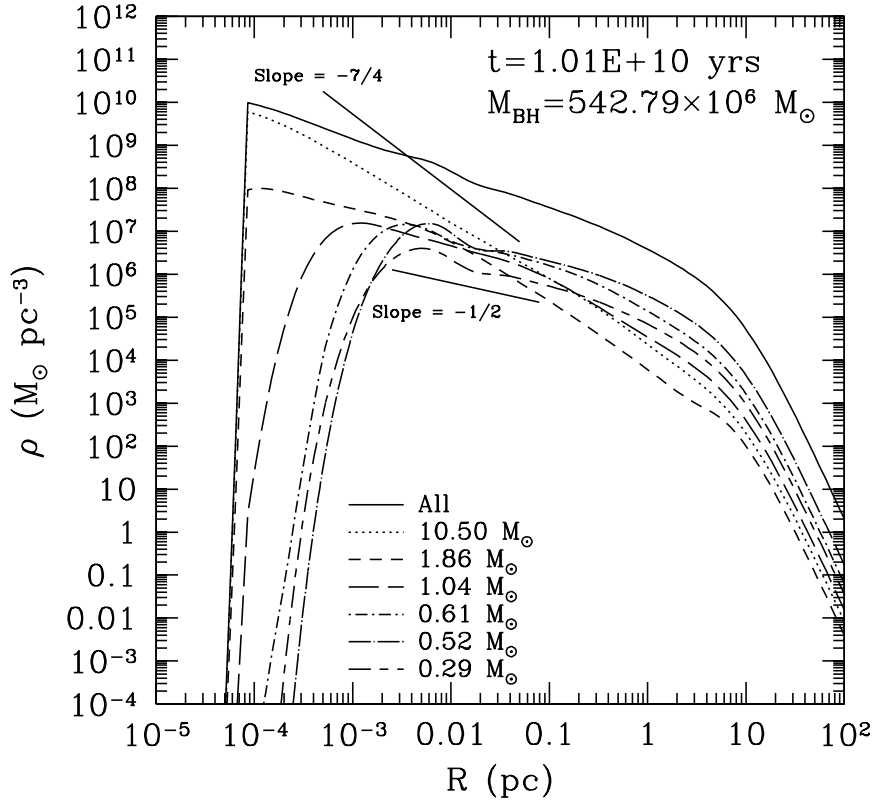}
\vspace{-1.2cm}
\caption{Stellar density profiles of a nucleus with mass 100 times
  that of the Galactic Center.}  
\end{figure}

\begin{figure}[!ht]
\epsscale{0.90}
\plotone{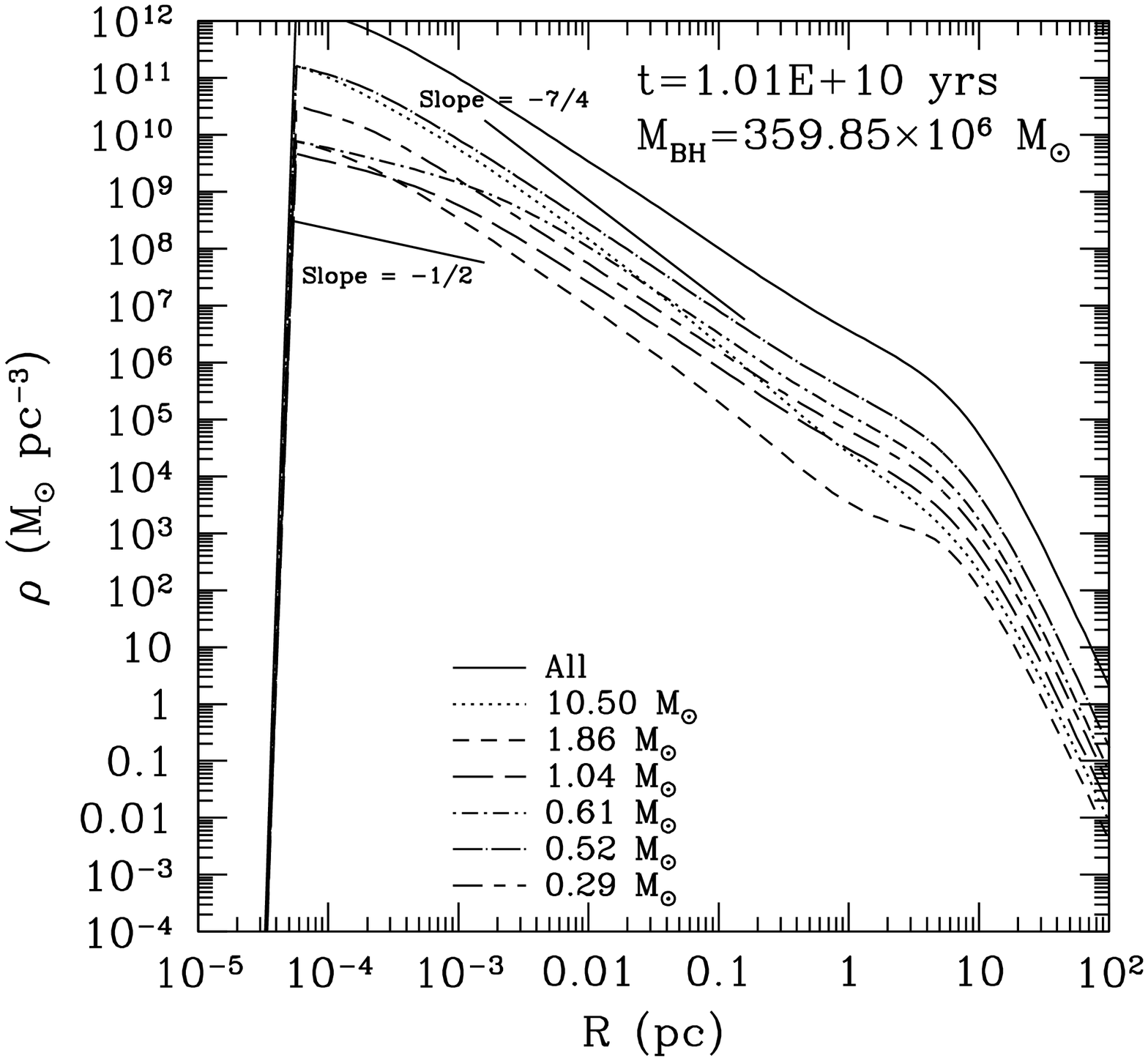}
\vspace{-1.2cm}
\caption{The same as Figure 6 but \textsl{not} including stellar collisions.}
\end{figure}

The stellar system in the Galactic Nucleus is most likely dynamically
relaxed and nearly spherically symmetric.  Using these assumptions it
is possible to model the mass distribution of differing mass groups
such as stellar-mass black holes, evolved giant stars, and main
sequence stars.  For this model, and the models shown in the previous
section I have used Fokker-Planck simulations similar to those
presented in Murphy, Cohn, \& Durisen (1991). Those models presented a
set of generic nuclei with varying mass function slopes and nucleus
masses.  The primary focus of the Murphy et al.~models was active
galactic nuclei and quasars.  Here I present a realistic model of the
Galactic Center.  A Kroupa (2001) mass function is used with initial
stellar masses ranging from 0.1 to 40 $M_{\sun}$.  The stellar system
starts with a seed central black hole of 700 $M_{\sun}$ and is allowed
to dynamically evolve. The stars are also allowed to evolve using the
prescription of Hurley, Pols, \& Tout (2000). The central seed black
hole is fed mass from the surrounding stellar population due to tidal
disruptions of stars, stellar collisions, and stellar evolution. All
mass loss due to stellar collisions and tidal disruptions is fed to
the central black hole due to the proximity of these events to the
black hole. A preset fraction of mass loss due to stellar evolution is
allowed to be fed to the central black hole and the remainder is
assumed to be ejected from the nucleus.  This preset fraction is a
model parameter that can be varied from 0 to 1.

In Figure 4 a model for the Galactic Nucleus at a Hubble time is
shown. By this time the central black hole has reached a mass of
4.4$\times$10$^6$ $M_{\sun}$.  Most of this growth was fueled by mass
loss from evolving stars, with less than 20$\%$ being due to tidal
disruptions of stars, and nearly nothing from collisions.  As would be
expected mass segregation has caused stellar-mass black holes to
dominate the stellar density distribution.  These black holes
have a power-law density slope of -7/4 while less massive stars have
progressively smaller slopes that approach -3/2 (Bachall \& Wolf 1977;
Murphy et al.~1991; and Freitag, Amaro-Seoane, \& Kalogera 2006).
Even though the central massive black hole dominates the gravitational
potential of the inner few parsecs the extended distribution of
stellar-mass black holes will play a significant role in the dynamics
of the observed giant stars.  Their presence could be found via
retrograde precession of observed stars.  It may also be possible to
observe scattering events due to this unseen population of massive
remnants. Inside of 10$^{-3}$ pc the densities and velocities are high
enough so that collisions begin to dominate and destroy stars on
orbits bound to the central black hole.  It should be noted that this
is a statistical representation of the stars.  In reality less than 50
$M_{\sun}$ lie within this radius, most of which would be stellar-mass
black holes.

\subsection{The Depletion of Late-Type Giant Stars}

Recent observations show what appears to be a depletion of late-type
giants in the Galactic Center (Do et al.~2009 and Buchholz,
Sch\"{o}del, \& Eckart 2009).  One common misconception is that this
is due to collisions.  But because this phase is relatively short
giants are unlikely to undergo a collision during post-main-sequence
evolution.  A more careful investigation of the Galactic Nucleus model
presented in the previous section indicates that for a Kroupa mass
function collisions of late-type giants \textsl{cannot} account for their
depletion (Geiss et al.~2010).  In fact collisions of giants are
insignificant in our Galactic Nucleus model.  The results indicate
that tidal disruptions of giants play a much more important role in
the resulting observed cusp, or lack thereof.  Our results show that
for the late-type giants the projected slope ranges from 0 for the
inner 1\arcsec~to a slope of -0.4 at 10\arcsec.  This flattening is
solely due to the tidal disruptions of giants and not collisions.
Another possible alternative has been mentioned by Davies in these
proceedings. It may be possible to account this depletion of giants by
assuming it is due to a top heavy mass function, one in which the
number of stellar-mass black holes is much larger than what a Kroupa
mass function would produce.  In this top-heavy scenario main sequence
stars would be destroyed before evolving into giants by the large
excess of stellar-mass black hole (Bartko et al.~2010).  Because
giants would be preferentially destroyed in the inner regions of
the nucleus a flat profile would presumably be produced. If this is
the case the relaxed population of main sequence stars should have a
density cusp with a -1/2 slope per section 2.3.  Unfortunately
confirmation of such a slope of the main sequence stars cannot be made
with present-day technology.

\subsection{More Massive Galactic Nuclei}

Besides the Galactic Center many other galactic nuclei are believed to
contain massive black holes.  The Andromeda Galaxy is believed to
harbor a black hole of $3\times10^{7}$ $M_{\sun}$, and M87 may have a
black hole of $3\times10^{9}$ $M_{\sun}$.  As the mass of the nuclear
cluster increases so does the amount of matter available to grow the
black hole.  And depending on the mass of the galaxy, collisions can
play and important role in its growth and the observed stellar
cusp. Figures 5 and 6 present the stellar density profiles of nuclei
with 10 and 100 times the mass of our own nucleus, respectively.  As
would be expected the model ten times more massive than the Galaxy
produces a black hole roughly 10 times more massive than the black
hole at the Galactic Center.  Perhaps the most representative example
of this model nucleus would be the Andromeda Galaxy.  It is also
interesting to note that though still small, collisions of stars are
playing a more important role in the model shown in Figure 5.  In
Figure 6 collisions dominate over tidal disruptions.  This is easily
seen in the slope of the density profiles. Though the black holes are
close to a -7/4 slope the main sequence stars have a power-law slope of
-1/2 out to 1 pc.  This slope is indicative of collisions depleting
these stars. Compare this to the identical model shown in Figure 7
which does not include stellar collisions.  One other item to note is
that more massive nuclei will actually have fewer stellar-mass black
holes interior to 0.01 pc.

\section{Conclusion}

Because stellar evolution and tidal disruptions of stars are the
predominant mass-loss mechanisms; the relaxed stellar system
at the Galactic Center should have a cusp with a density power-law
slope of -7/4 for stellar-mass black holes to -3/2 for main sequence
stars.  But for the late-type giant stars the slope will be
significantly less than this, due not to collisions, but due to tidal
disruptions of these large stars.  For massive nuclei stellar
collisions play a much more important role in the development of a
stellar cusps.  Because collisions dominate these more massive nuclei
their cusps will have a -1/2 power-law slope.

\acknowledgements  The assistance of K. Phifer, B. Geiss,
M. McFall, and the Butler Institute for Research \& Scholarship with
this research is appreciated.

\end{document}